\documentclass[nameyear]{elsart}
\usepackage{epsfig}
\usepackage{amssymb}

\newcommand{\msun}{\mbox{$M_\odot$}}
\newcommand{\rsun}{\mbox{$R_\odot$}}
\def\be{\begin{eqnarray}}
\def\ee{\end{eqnarray}}
\def\lsim{\mathrel{\rlap{\lower3pt\hbox{\hskip1pt$\sim$}}
     \raise1pt\hbox{$<$}}} 
\def\gsim{\mathrel{\rlap{\lower3pt\hbox{\hskip1pt$\sim$}}
     \raise1pt\hbox{$>$}}} 

\def\COrate{$^{12}$C($\alpha,\gamma$)$^{16}$O\ }

\begin{document}

\runauthor{Bethe, Brown, \& Lee}

\begin{frontmatter}
\title{Evolution and Merging of Binaries with Compact Objects} 

\author[cornell]{Hans A. Bethe\corauthref{bethe}}
\author[suny]{Gerald E. Brown} 
\author[pnu]{Chang-Hwan Lee} 

\address[cornell]{Floyd R. Newman Laboratory of Nuclear Studies,
Cornell University, \\ Ithaca, NY 14853}

\address[suny]{Department of Physics and Astronomy,
               State University of New York,\\ Stony Brook, NY 11794, USA}

\address[pnu]{
Department of Physics, Pusan National University,
              Busan 609-735, Korea\\
{\small (E-mail: clee@pusan.ac.kr)}}

\corauth[bethe]{Deceased (July 2, 1906 $-$ March 6, 2005).}


\begin{abstract}
In the light of recent observations in which short $\gamma$-ray
bursts are interpreted as arising from black-hole(BH),
neutron-star(NS) or NS-NS mergings we would like to review our
research on the evolution of compact binaries, especially those
containing NS's. These were carried out with predictions for LIGO in
mind, but are directly applicable to short $\gamma$-ray bursts in
the interpretation above.

Most important in our review is that we show that the standard
scenario for evolving NS-NS binaries always ends up with a low-mass
BH (LMBH), NS binary. Bethe and Brown (1998) showed that this fate
could be avoided if the two giants in the progenitor binary burned
He at the same time, and that in this way the binary could avoid the
common envelope evolution of the NS with red giant companion which
sends the first born NS into a BH in the standard scenario. The
burning of He at the same time requires, for the more massive giants
such as the progenitors of the Hulse-Taylor binary NS that the two
giants be within 4\% of each other in ZAMS mass. Applying this
criterion to all binaries results in a factor $\sim 5$ of LMBH-NS
binaries as compared with NS-NS binaries.

Although this factor is substantially less than the originally
claimed factor of 20 which Bethe and Brown (1998) estimated, largely
because a careful evolution has been carried through here, our
factor 5 is augmented by a factor of $\sim 8$ arising from the
higher rate of star formation in the earlier Galaxy from which the
BH-NS binaries came from. Furthermore, here we calculate the mergers
for short-hard gamma-ray bursts, whereas Bethe and Brown's factor 20
included a factor of 2 for the higher chirp masses  in a BH-NS
binary as compared with NS-NS one. In short, we end up with an
estimate of factor $\sim 40$ over that calculated with NS-NS binary
mergers in our Galaxy alone. Our total rate is estimated to be about
one merging of compact objects per year.

Our scenario of NS-NS binaries as having been preceded by a double
He-star binary is collecting observational support in terms of the
nearly equal NS masses within a given close binary.

We review our work on population synthesis of compact binaries,
pointing out that it is in excellent agreement with the much more
detailed synthesis carried out by Portegies Zwart. This is currently
of interest because the recent discovery of the double pulsar has
substantially increased the number of binary NS's that will merge
gravitationally, giving signals to LIGO. This discovery brings in
the low ZAMS mass main sequence progenitors that can evolve into a
NS binary, adding importantly to the ``visible" binaries that can
merge. However it does not affect the factor $\gsim 40$ increase,
mostly from the much greater number of LMBH-NS binaries, which have
only a small probability of being observed before they merge.

We develop the phenomenology which suggests that NS's evolve from
ZAMS mass $\sim 10-18 \msun$ star, LMBH's from $18-20 \msun$, and
high-mass BH's from $20-30\msun$. These brackets follow from
Woosley's \COrate rate of 170 MeV barns at 300 keV.

We discuss the observed violation of our previous maximum NS mass
$M_{\rm NS}^{\rm max}=1.5\msun$, raising our $M_{\rm NS}^{\rm max}$
to $1.7\msun$ and comment on how our scenario would change if the
maximum NS mass is greater than $1.7\msun$.

\end{abstract}

\end{frontmatter}


\section{Introduction\label{intro}}

It is now 27 years since Bethe et al. 1979 
(denoted as BBAL) was published. At that time we thought the
supernova problem to be solved. Although a tremendous amount of
work, much of it numerical, has gone on and even more is going on
now, the better the physics input, the further the explosion is from
success. However, see the paper by Adam Burrows in this volume.

This sorry situation has led to all sorts of opinions about the nature
and evolution of compact objects, the authors feeling free to extrapolate
in all directions, since the basic mechanism for producing them does
not work. It is not generally known, or if it is known, not generally
believed that there is considerable phenomenology developed which
connects different phenomena. This was developed chiefly by
Woosley and his students in the study of numerical NS
formation. On the evolutionary side and the connection of various
types of compact stars, much of it was developed by van den Heuvel
and his students.

We used their techniques to make many connections between compact stars
in the ``Formation and Evolution of Black Holes in the Galaxy"
(Bethe, Brown, \& Lee 2003).
In this paper we wish to expose in a simple way and summarize some
of the connections. At this stage our connections mostly have to be
considered in a pragmatic sense; they are as good as they work.
We hope that we show that they work well, in that one can understand
a lot through them.

We adopt the method of population synthesis in evolving binaries
with compact companions (Bethe and Brown 1998).
We update this work in terms of the much more extensive and more
accurate work of Belczynski et al. (denoted as BKB) (2002).
These authors performed detailed parameter studies.
Many more investigators have tried to estimate the number of binaries
and their mergers from those we observe, and then extrapolating
to the entire Galaxy and then from our Galaxy to many other
galaxies.
There are two main problems with this latter
method. (i) It is not easy to see systems that emit radio emission
only weakly; large corrections have to be made for those we don't see.
This difficulty will become clear with our discussion in Sec.~\ref{sec3}
of the newly discovered double pulsar which increases the number of
observable gravitational mergers estimated from observed systems
by at least an order of magnitude (Kalogera et al. 2004).
(ii) Mergers of BH-NS binaries are much more probable than mergers
of binary NS's. Indeed, the signal from the former will tend to be
greater because of the larger chirp mass implied by the BH mass
being greater than the NS mass. Yet there is of yet little
probability of seeing the BH-NS binaries, the number of which we
estimate to be 5 times greater than binary NSs, because the latter
are observable for $\sim 100$ times longer than the former, due to
the fact that the magnetic field of the first born NS (which turns
into a BH in the BH-NS binaries) in a double NS binary gets
recycled, by mass accreted from its companion, which brings its
magnetic field down a factor $\sim 100$, and, as we outline,
increases the time it can be observed by about the same factor (See
our later discussion).

In population synthesis one estimates the frequency of supernova explosions
in our Galaxy from those in similar spiral galaxies.
(Many in our Galaxy are thought to be obscured by the milky way.)
Then from an estimate of binarity, say 50\%, one has the number of
binaries in which both stars are massive enough to go supernova.
In fact, the calculations of Bethe and Brown (1998) proceeded in
parallel with those of Portegies Zwart and Yungelson (1998). The
latter assumed a Galactic supernova rate of $0.015$ yr$^{-1}$, Bethe
and Brown (1998), $0.0225$ yr$^{-1}$.


\section{Connection of Fe core and compact star masses}
\label{sec2}

Most troublesome in the study of compact stars is the lack of
connection between the Fe core, which can be and has been
calculated, most recently by Alex Heger (Brown et al. 2001a)
with Woosley's Kepler, and the mass of the compact object. Woosley
chooses the outer edge of the Fe core to be at the location of
a large discontinuous change in $Y_e$ which marked the outer
extent of the last stage of convective silicon shell burning.

Bethe and Pizzochero (1990)
used for SN 1987A a schematic but realistic treatment of the radiative
transfer problem, which allowed them to follow the position in mass of
the photosphere as a function of time. They showed that the observations
determine uniquely the kinetic energy of the envelope once its mass is
known. They obtained a kinetic energy of the ejecta $\ge$ 1 foe
($=10^{51}$ erg), the energy scaling linearly with $M_{\rm env}$.
The main point is that this was done without any input from the numerical
models used to describe 1987A. From the envelope masses considered,
the range of energies was $\sim 1-1.4$ foe.

Using the fact that following the supernova explosion as the shock moves
outwards the pressure is mainly in radiation, Bethe and Brown (1995)
showed from the known value of $\sim 0.075\msun$ of $^{56}$Ni production
in 1987A that
an upper limit on the gravitational mass of $\sim 1.56 \msun$ could
be obtained for the progenitor.
The main point was that the matter is very dilute in the bifurcation
region so that the amount of fallback depends only weakly on the precise
separation
distance chosen, also that the amount of fallback is roughly equal in
magnitude to the binding energy of the compact object. Thus, the Fe core
mass is a good estimate of the mass of the compact object. (See also
Table~3 of Brown, Weingartner \& Wijers (1996) where the amounts of
fallback material from distances of 3500 and 4500 km determined by
Woosley are plotted.)

Our conclusion is that we can use calculated Fe core masses as
an estimate for the masses of the compact cores which will result.


Woosley outlined in several publications the important role of
the magnitude of the \COrate reaction in determining the
ZAMS (zero age main sequence)
mass at which stars would go into BHs.
Whereas $^{12}$C is formed essentially by the triple $\alpha$-reaction
which goes as the square of the density, the reaction \COrate goes
with the density, being a binary reaction. With increasing ZAMS mass,
the central density of stars decreases, the mass going roughly as the
square root of the radius. Thus, there comes a ZAMS mass at which the
carbon is removed by the \COrate reaction as fast as it is formed.
At that mass, there is essentially no $^{12}$C to be burned.

Now as long as $^{12}$C has to be burned, it does so at a temperature
of $\sim 80$ keV, $\sim 4$ times greater than that at which the
\COrate removes carbon. In burning at such a high temperature a lot of
entropy is carried away, decreasing the entropy substantially.

BBAL (Bethe et al. 1979)
showed that the final entropy per nucleon is $\sim 1$ (in units where
$k_B=1$) so that the way in which the higher entropy achieved when
convective carbon burning is skipped, is that the Fe core increases
substantially in mass. This is just the mass at which stars begin evolving
into BHs. A more complete argument of this is given in
Brown et al. (2001a).

\begin{figure}
\centerline{\epsfig{file=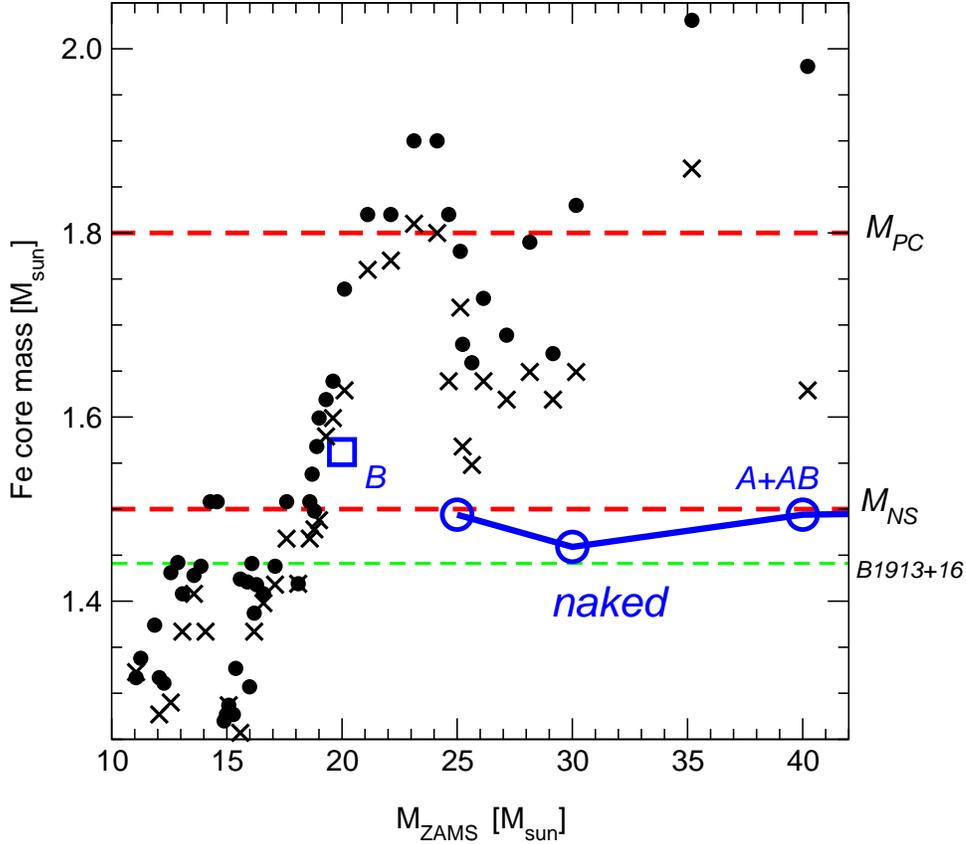,height=14cm}}
\caption{Reproduction of Fig.~2 of Brown et al. (2001a).
Comparison of the iron core masses resulting from
the evolution of ``clothed" and ``naked" He cores.
Filled circles and crosses correspond to the core masses of ``clothed"
stars at the time of iron core implosion for a finely spaced grid of
stellar masses (Heger, Woosley, Martinez-Pinedo, \& Langanke 2001).
The circular black dots were calculated with the Woosley \& Weaver 1995 code,
whereas the crosses employ the vastly improved Langanke,
Martinez-Pinedo (2000) rates for electron capture and beta decay.
Open circles (square) correspond to the naked He stars in case A$+$AB
(B) mass transfer of Fryer et al. (2002), with reduced WR
mass loss rate.  If the assembled core mass is
greater than $M_{\rm PC}= 1.8\msun$, where $M_{PC}$ is the
proto-compact star mass as defined by Brown \& Bethe (1994), there is
no stability and no bounce; the core collapses into a high mass BH.
$M_{NS}=1.5\msun$ denotes the maximum mass of NS
(Brown \& Bethe 1994).  The mass of the heaviest known well-measured
pulsar, PSR B1913$+$16, is also indicated with dashed horizontal line
(Thorsett \& Chakrabarty 1999).}\label{fig1}
\end{figure}

Alexander Heger, using Woosley and Weaver's carbon burning rate of
170 keV barns, and the best current physics, reevolved the main sequence
stars with the results shown in Fig.~\ref{fig1}.
The input physics and the results given here were summarized in Heger
et al. (2001).
The experimental determination of the carbon burning rate is
the Stuttgart  one (Kunz et al. 2001):
\be
S_{\rm tot}^{300} = (165\pm 50)\ {\rm keV\ barns}.
\ee
A recent paper summarizing all of the Stuttgart work to date (Hammer
et al. 2005), which also consider the data from elastic scattering
and from the decay of $^{16}$N, adds up to:
\be
S_{\rm tot}^{300} = (162\pm 39)\ {\rm keV\ barns}.
\ee
The $^{12}$C($\alpha,\gamma$) experiment was so time-consuming that
it required far more time than the usual 3-5 years generally devoted
to an experiment. For the project, the Stuttgart team spent a
total of 262 days of beam time, not counting all the days of preparation.
In the Stuttgart experiment all up to date technical achievements
were combined into a single experiment.

One of the authors of the present paper (G.E.B.) tried a number of
times to make a shell-model calculation of this rate. But the
unperturbed one-particle, one-hole state and (deformed)
three-particle, three hole state mix destructively, giving a small
net contribution to the 7.12 MeV 1$^{-}$ state in $^{16}$O, so the
matrix element to this state, which then decays to the $^{16}$O
ground state by emitting the $\gamma$-ray, could not be accurately
calculated.

Willy Fowler once said that no nuclear reaction that can be measured
in the laboratory should be determined by astronomical observations.
However, Woosley's 170 keV barns, near the central value of the
Stuttgart measurements, fits our phenomenology so well,
especially the LMBH in SN1987A which had progenitor
ZAMS mass of $18-20\msun$, combines the laboratory measurement
and astronomical phenomenology. To within the accuracy of the
former, we believe the question to be settled.

The much used compilation by Schaller et al. (1992) uses an S-factor
of $\sim 100 $ keV barns.  They bring their central abundance of He
down to 0.16 at the end of core He burning only for a $25\msun$
star, below the point for convective He burning, so we believe that they
would start evolving high-mass BHs at this mass, $\sim 5\msun$
higher than with the Woosley rate of 170 keV barns.

{}From Fig.~\ref{fig1} we see that the Fe cores, which we identify with the
final compact objects, as outlined in the last section, increase rapidly
in mass
in the region around $20\msun$ going above our $1.5\msun$ maximum NS
mass at $\sim 18\msun$, just the ZAMS mass of 1987A which we believe
to have gone into a LMBH.
We reiterate here that we have estimated the compact object mass to be
the same as that of the Fe core, the fallback in the latter case compensating
for the additional gravitational attraction in the former case.
Then at $\sim 20\msun$ the Fe
cores climb above $1.7-1.8\msun$ which is our limit for high-mass BHs;
i.e., those in which the He envelope is not exploded outwards
as in 1987A, but collapses inwards. In Lee, Brown, and Wijers (2002) we
find the high-mass BHs in the Galaxy can be made from
$20-30\msun$ stars. Even though the Fe core masses come down somewhat
above $23\msun$, the envelopes are so massive that they will collapse
inwards.

We believe that the key to the different regions in which NS's,
LMBHs (in the very narrow region of $18-20\msun$) and high-mass BHs
can be evolved is the \COrate rate of $\sim 170$ keV barns
introduced by Woosley. In any case, taking the Fe core masses to be
those of the compact object, we do well on the phenomenology of the
different regions of ZAMS masses which give NS's and LMBHs. Whereas
a large number of high-mass BHs have been found by now, 17 in the
transient sources (Lee, Brown, Wijers 2002) no LMBH in a binary has
been identified. But the possible range of $\sim 18 -20 \msun$ is
very narrow, with only the progenitor of SN 1987A from that range.

\section{Evolution of Binary Neutron Stars}
\label{sec3}

In Table~\ref{tab1} is the compilation by Lattimer and Prakash (2004) of the
compact objects in binaries.

\begin{table}
\caption{Compilation of the compact objects in binaries by Lattimer
and Prakash (2004). References are given in their paper.
$^\star$Brown et al. (1996) argue that 4U~1700$-$37, which does not
pulse, is a LMBH. $^\dagger$We have added, following the comma, the
recent measurement of Van der Meer et al. (2004).
$^{\dagger\dagger}$Results for J0751$+$1807 is from Nice et al.
(2005). } \label{tab1}
\begin{center}
\begin{tabular}{llll}
\hline
Object      & Mass ($\msun$) \phantom{xxxxxx} & Object    & Mass ($\msun$) \\
\hline
\multicolumn{4}{l}{\it X-ray Binaries} \\
4U1700$-$37$^\star$ & 2.44$^{+0.27}_{-0.27}$ &
Vela X-1    & 1.86$^{+0.16}_{-0.16}$\\
Cyg X-1     & 1.78$^{+0.23}_{-0.23}$ &
4U1538$-$52 & 0.96$^{+0.19}_{-0.16}$ \\
SMC X-1$^\dagger$     & 1.17$^{+0.16}_{-0.16}$, 1.05$\pm$0.09 &
XTE J2123$-$058 & 1.53$^{+0.30}_{-0.42}$ \\
LMC X-4$^\dagger$     & 1.47$^{+0.22}_{-0.19}$, 1.31$\pm$0.14 &
Her X-1     & 1.47$^{+0.12}_{-0.18}$ \\
Cen X-3$^\dagger$     & 1.09$^{+0.30}_{-0.26}$, 1.24$\pm$0.24 &
2A 1822$-$371   & $> 0.73$ \\ 
\multicolumn{4}{l}{\it Neutron Star - Neutron Star Binaries} \\
1518$+$49           & 1.56$^{+0.13}_{-0.44}$ &
1518$+$49 companion & 1.05$^{+0.45}_{-0.11}$\\
1534$+$12           & 1.3332$^{+0.0010}_{-0.0010}$ &
1534$+$12 companion & 1.3452$^{+0.0010}_{-0.0010}$ \\
1913$+$16           & 1.4408$^{+0.0003}_{-0.0003}$ &
1913$+$16 companion & 1.3873$^{+0.0003}_{-0.0003}$ \\
2127$+$11C           & 1.349$^{+0.040}_{-0.040}$ &
2127$+$11C companion & 1.363$^{+0.040}_{-0.040}$ \\
J0737$-$3039A        & 1.337$^{+0.005}_{-0.005}$ &
J0737$-$3039B        & 1.250$^{+0.005}_{-0.005}$ \\
J1756$-$2251         & 1.40$^{+0.02}_{-0.03}$ &
J1756$-$2251 companion & 1.18$^{+0.03}_{-0.02}$ \\
\multicolumn{4}{l}{\it Neutron Star - White Dwarf Binaries} \\
B2303$+$46  & 1.38$^{+0.06}_{-0.10}$ &
J1012$+$5307 & 1.68$^{+0.22}_{-0.22}$ \\
J1713$+$0747 & 1.54$^{+0.007}_{-0.008}$ &
B1802$-$07   & 1.26$^{+0.08}_{-0.17}$ \\
B1855$+$09   & 1.57$^{+0.12}_{-0.11}$ &
J0621$+$1002 & 1.70$^{+0.32}_{-0.29}$ \\
J0751$+$1807$^{\dagger\dagger}$ & 2.10$^{+0.20}_{-0.20}$ &
J0437$-$4715 & 1.58$^{+0.18}_{-0.18}$ \\
J1141$-$6545 & 1.30$^{+0.02}_{-0.02}$ &
J1045$-$4509 & $<$ 1.48 \\
J1804$-$2718 & $<$ 1.70 &
J2019$+$2425 & $<$ 1.51 \\
\multicolumn{4}{l}{\it Neutron Star - Main Sequence Binaries} \\
J0045$-$7319 & 1.58$^{+0.34}_{-0.34}$ \\
\hline
\end{tabular}
\end{center}
\end{table}

When we say ``evolution" we do not mean a complete one with calculation
of kick velocities, etc. in NS formation. Rather, we chiefly
discuss the difference in masses of the pulsar and its companion.

Such a situation occurs in the scenario for making binary pulsars,
in which the spiral-in of the NS through the companion supergiant
expels the envelope, which is hydrodynamically coupled to the drag
from the NS, leaving a He star as companion. Hypercritical accretion
rates are encountered and Chevalier worked out that for $\dot M
\gsim 10^4\ \dot M_{\rm Edd}$ radiation pressure is unable to limit
the accretion the photons are simply carried in by the adiabatic
inflow.\footnote{In fact, this happens at a much lower rate, but the
$10^4$ Eddington limit is needed to build up sufficient density in
the accretion shock so that the energy can be carried off by
neutrinos.} Chevalier estimated that sufficient mass would be
accreted by the NS during the common envelope evolution to send it
into a BH. Said more simply for such a high $\dot M$ the drift
(random walk) velocity is inwards onto the NS.

Although not generally accepted by astronomers, partly because it
destroyed the usual scenario for binary pulsar evolution, the
trapping of neutrinos and their being carried in by the adiabatic inflow
in the collapse of large stars as in BBAL\cite{BBAL}
involved the same mechanism, but of course with different parameters.
Without the pressure from trapped neutrinos, supernova explosions wouldn't
have any chance of succeeding.

The Brown scenario was worked out in detail by Bethe and Brown
(1998) who showed that the first born NS in the conventional
scenario would accrete $\sim 1\msun$ in the about one year of common
envelope evolution, taking it into a BH. A simple analytical
treatment was given of the common envelope evolution.


Belczynski et al. (2002) removed the approximation of neglecting the
compact object mass made by Bethe and Brown, so that the common
envelope problem could be accurately handled by solving a series of
partial differential equations. This lowered the Bethe and Brown
accretion by about 25\%. In the rest of this paper we shall use the
Belczynski et al. way.

The Hulse-Taylor pulsar 1913$+$16 of mass $1.442 \msun$ with
companion mass $1.387 \msun$ is the most massive of the NS's in
double NS binaries. Burrows and Woosley (1986) evolved it from ZAMS
$\sim 20\msun$ giants. It can be seen from Fig.~\ref{fig1} that this
is just where the Fe core masses increase rapidly with change in
mass, possibly giving an explanation of why the difference in pulsar
and companion masses is nearly 4\%, close to our upper limit for
overlapping He burning.

Going down in NS mass we reach 1534$+$12 and 2127$+$11C and its
companion although the latter is in a globular cluster and often
said to have been formed only later by exchange of other stars in
binaries. In both of these binaries the two masses are within $\sim
1\%$ of each other. Note that in 1534 the companion mass is greater
than that of the pulsar. It can be seen from Fig.~\ref{fig1} that
right around ZAMS $15\msun$, there are fluctuations where the Fe
core mass from the more massive ZAMS mass is lighter than from the
less massive one. In any case, we believe that 1534 and 2127 come
from ZAMS masses more or less in the middle of our range from
$10-20\msun$. A ZAMS mass $15\msun$ corresponds to a He core of
$\sim 4\msun$, so this is approximately the limit below which the He
stars evolve in red giant.

At the bottom of our interval comes the double pulsar J0737A, B and
1756$-$2251, which we take to have evolved from giants of ZAMS mass
$10-12\msun$. In both of these the pulsars will have accreted some
mass in the He-star red giant evolution, so called Case BB mass
transfer. As we discuss later, this mass transfer is difficult to
calculate quantitatively, because He-star winds are large and
unpredictable. We assign an uncertainty of $\lsim 0.2\msun$ to the
amount of He accreted in Case BB mass transfer, which we return to
later in our discussion.

We believe that we have well-measured binaries at the top and bottom
end of our interval ZAMS $10-20\msun$ and two well-measured binaries
1534 and 2127 in the middle.


Using evolutionary calculations of S.E. Woosley of He stars of mass
$<4\msun$ (roughly corresponding to $15\msun$ ZAMS) in which wind
loss is included, Fryer and Kalogera (1997) show that only special
conditions in terms of kick velocities allow the pulsar in close
NS-He binaries to avoid the envelope of these low-mass He stars. In
detail, Fryer and Kalogera (1997) found that if they worked
backwards from the system of 1913$+$16 and assume there is no kick,
then the preexplosion separation is less than the Roche lobe radius
of the helium-star, NS system, leading to a pulsar, helium-star
common envelope. Because in the standard scenario (not double
helium-star) scenario of binary pulsar evolution, the first mass
exchange of the two giant progenitors is usually during the red
giant phase (Case B mass transfer) of the more massive progenitor,
then the later red giant evolution of the helium star in the
helium-star, NS phase, which implies a second common envelope, is
called in the literature Case BB mass transfer.

In any case, from Fig.~\ref{fig1} with our approximation of the Fe
core mass plus fallback being equal to the NS mass, we see that the
NS's in 1534$+$12 probably came from the middle of the range $\sim
10-20 \msun$.

The small $\sim 1\%$ difference in masses in 1534$+$12 looks like a
striking confirmation of the double He-star scenario. The companion
is $\sim 1\%$ more massive than the pulsar. To bring its magnetic
field down to $10^{12}$ G we estimate that with the Eddington rate
and He burning time possibly as large as double that for the
companion in the Hulse-Taylor pulsar, then the pulsar would accrete
$\sim 0.03 f\msun$. We have motivated (Francischelli et al. 2002)
$f\sim 0.1$ from the propeller effect, so this need not be much.

We put off a discussion of the evolution of the double pulsar
J0737$-$3039 and of J1756$-$2251 where the pulsar must have accreted
matter during the helium star red giant phase of the companion, Case
BB mass transfer, until after we have discussed the 4 more massive
binaries.

We now discuss the standard scenario for binary NS formation (van
den Heuvel and van Paradijs 1993) and why it does not work. The
Bethe and Brown (1998) work was analytical, and the approximation of
neglecting the mass of the compact object in comparison with the
companion, while in transition from main sequence to He star, mass
was made. As noted above, BKB (2002) removed this approximation in
their numerical calculations and we now adopt their method.


\begin{table}
\begin{center}
\caption{Flat Distribution: Proximity probability to evolve NS
binaries. $M_1$ is the giant mass, $\Delta M$ is the 4\% mass
difference within which the binaries evolve into NS-NS binaries, and
$P$ is the probability of having mass difference within 4\%.}
\label{tabX1}

\vskip 4mm
\begin{tabular}{ccc}
\hline
$M_1$ [$\msun$] & $\Delta M=0.04\ M_1$ [$\msun$] & $P=\Delta M /(M_1 -10\msun)$ \\
\hline
20 & 0.80 & 0.08 \\
19 & 0.76 & 0.08 \\
18 & 0.72 & 0.09\\
17 & 0.68 & 0.10 \\
16 & 0.64 & 0.11 \\
15 &  0.60 & 0.12 \\
14 & 0.56 & 0.14 \\
13 & 0.52 & 0.17 \\
12 & 0.48 & 0.24 \\
11 & 0.44 & 0.44 \\
10 & 0.40 & $-$\\ \hline
\end{tabular}
\end{center}
\end{table}

We follow Pinsonneault and Stanek (2006) in evolving NS binaries,
but require the two massive progenitors to be within $4\%$ in mass
in order to burn He at the same time, rather than the 5\% they use.
Using a flat distribution and the fact that the IMF for the second
star is not independent of the first star, because it must be of
lower mass to evolve later than the first star, one finds that NS
binaries should be formed 16\% of the time, but 44\% of the time if
$M_1=11\msun$. That's where the twin is most likely formed. We show
this in Table~\ref{tabX1}.

Pinsonneault and Stanek (2006) assembled evidence that ``Binaries
like to be Twins". They showed that a recently published sample of
21 detached eclipsing binaries in the Small Magellanic Cloud can be
evolved in terms of a  flat mass function containing 55\% of the
systems and a ``twins" population with $q> 0.95$ containing the
remainder. All of the binaries had orbital period $P< 5$ days, with
primary masses $6.9 \msun <M_1 <27.3 \msun$.

Historically large selection effects have been identified (Goldberg
et al. 2003; Hogeveen 1992). These will lower the number of twins
found by Pinsonneault and Stanek.

The important role of twins is that the two giants are close enough
in mass \footnote{Pinsonneault and Stanek used 5\% whereas we prefer
4\% as will be discussed.} that in Brown's (1995) scenario they can
evolve into NS-NS binaries, whereas if they are further apart in
mass they will evolve into a LMBH-NS binary (Chevalier 1993; Bethe
and Brown 1998).

Thus the twins may increase the number of NS-NS binaries. We suggest
that the resulting number of short hard gamma-ray bursts, which
result from the merging of the binaries, which to date are unable to
differentiate between the two species, may not be changed much, some
of the predicted large excess of LMBH-NS binaries appearing rather
as NS-NS binaries. However, because the latter are so much more
easier to observe, the role between what we see and what is present
will be tightened.

We point out that Belczy\'nski et al. (2002) in their simulation D2
in which the maximum NS mass is $1.5\msun$ and the mass proximity in
the progenitor binaries (to evolve NS's) is taken, like Pinsonneault
\& Stanek to be 5\%, obtain a ratio of 4 for (BH$+$NS)/(NS$+$NS) and
would obtain the ratio of 5 had they used our 4\% proximity in
masses.

In their Case D2 BKB (2002) find a total gravitational merging rate
of $0.45 \times 10^{-4}$ yr$^{-1}$ for the sum of their double NS
and BH-NS mergings to compare with the Bethe \& Brown $0.70 \times
10^{-4}$ yr$^{-1}$ once the Bethe \& Brown supernova rate of
$0.0225$ yr$^{-1}$ is lowered to the BKB $0.0172$ yr$^{-1}$.

In short, there is general agreement amongst the authors quoted
above, except that it is not clear how many twins will be left once
selection effects are taken into account. For simplicity we shall
use a total gravitational merging rate in our Galaxy of $10^{-4}$
yr$^{-1}$.

Whereas we call the standard scenario of binary NS evolution that of
van den Heuvel and van Paradijs (1993), BKB include hypercritical
accretion in what they call their standard scenario. We believe that
their case D2 with $M_{\rm NS}^{\rm max}=1.5\msun$ is strongly
favored by the closeness in mass of the double NS binaries.

The Bethe \& Brown (1998) work did not cover the Case BB mass
transfers in the binaries from the less massive ZAMS masses $\lsim
15\msun$, which we discuss in Sec.~\ref{sechyper}.

\section{Observability Premium}

The behavior of the pulsar magnetic field is crucial. Van den Heuvel
(1994b) has pointed out that NS's formed with strong magnetic fields
$10^{12} - 5\times 10^{12}$ G, spin down in a time
\be
\tau_{\rm sd} \sim 5\times 10^6\ {\rm yrs} \label{eq4}
\ee
and then disappear into the graveyard of NS's. (The pulsation
mechanism requires a minimum voltage from the polar cap, which can
be obtained from $B_{12}/P^2 \gsim 0.2$ with $B_{12}=B/10^{12}$ G
and $P$ in seconds.) The relativistic binary pulsar 1913$+$16 has a
weaker field $B\simeq 2.5\times 10^{10}$ G, and therefore emits less
energy in magnetic dipole radiation. Van den Heuvel estimates its
spin-down time as $10^8$ yrs. There is thus a premium in
observational time for lower magnetic fields, in that the pulsars
can be seen for longer times.

Wettig and Brown (1996) used van den Heuvel's idea to invent the
Observability Premium
\be
\Pi = \frac{10^{12} \rm G}{B} \label{eq5}
\ee
where $B$ is the magnetic field of the pulsar. $\Pi$ gives the time
relative to that of a $10^{12}$ G pulsar, that the pulsar can be
observed. Taam and van den Heuvel (1986) found empirically that the
magnetic field of a pulsar dropped roughly linearly with accreted
mass. Thus, the Observability Premium is high, given a large amount
of such mass.

Wettig and Brown (1996) brought the Observability Premium $\Pi$ into
the weighting in their evolution of binary pulsars, assuming because
of the high winds during He burning that accretion occurred only in
the NS, He-star stage. Since the maximal accretion of $\dot M =
3\times 10^{-8} \msun$ yr$^{-1}$ adds up to make the pulsar
observable for a longer time, this gave an explanation of the
relatively large number of narrow, short period binary pulsars. As
noted earlier, the actual accretion rate may be an order of
magnitude smaller because of the propeller effect (Francischelli et
al. 2002).

Whereas we have considered LMBH-NS binaries above, the formation of
NS binaries at redshift $z\sim 1$ due to the higher star formation
rate means that they should play a role increasing Kalogera et al.'s
rate. The Hulse-Taylor pulsar and 1534$+$12 have magnetic fields
$B\sim 10^{10}$ G so their observability premium; i.e., the longer
time that they can be observed, is $\sim 100$ times that of the Crab
pulsar, or $\sim 500$ megayears. In fact, the Hulse-Taylor pulsar is
estimated to merge in $\sim 300$ Myr. One would expect the same
observability premiums for NS binaries that follow from the $\gsim
15\msun$ giant progenitors to be similarly recycled and have similar
lifetimes because Wettig and Brown (1996) have pointed out that
their recycling occurs mostly in the He star, pulsar stage of their
formation, and the observational premium favors the formation of
binaries sufficiently close so that the accretion from the He-star
winds is at or near the Eddington limit. Thus, by a Gyr the pulsar
will have run down, and the binary will be invisible. Because of the
larger rate of star formation at redshift $z\sim 1$, there should be
$\sim 8$ times more of these than the binaries such as Hulse-Taylor
where one still sees the pulsar in operation. We shall develop this
point in more detail in the next section.

It is not clear that the matter accreted in the Case BB mass
transfer, that in the He red giant stage, as in the double pulsar
plays a similar role in bringing the magnetic field of the first
born pulsar down, but here $B=6.3 \times 10^9$ G so that the pulsar
would also run down and stop pulsing in Gyr. Thus, we would expect a
factor $\sim 8$ increase in Kalogera et al.'s rate from the double
NS binaries born some Gyr ago because of the higher star formation
rate.

\section{Short Hard Gamma-Ray Bursts}

The exciting new development which occurred while writing this
review is the observation of the short-hard gammy-ray bursts (SHBs),
made possible by the satellites Swift and Hete-2, which are able to
detect the $\sim 1$ second bursts and radio their positions to the
telescopes which can observe their afterglows. The four afterglows
of short-hard gamma-ray butsts and the progenitors inferred for
these bursts $-$ as well as for another four SHBs with known or
constrained redshift form the basis for the analysis of Nakar et al.
(2005) which we follow here. (See the Physics Reports by Nakar in
this volume.)

We will discuss these SHBs as chiefly  resulting from the LMBH-NS
binaries of Bethe and Brown (1998) since we believe that they
predominantly result from mergers of these binaries.

There is some evidence of beaming in the SHBs, as in the longer
GRBs. Two SHBs (050709 and 050724) have shown a steepening that can
be interpreted as a hint of a jet. This interpretation would
indicate a beaming factor of $\sim 50$ (Fox et al. 2005). Such a
beaming would reduce substantially the average isotropic energy and,
therefore, make it more difficult for LIGO to observe.

The Nakar et al. (2005) ``best guess" at SHBs is $R_{SHB}\approx
10^5$ Gpc$^{-3}$ yr$^{-1}$ with the assumption of beaming factor of
50. The rate of Bethe-Brown mergers in our Galaxy was estimated at
$\sim 10^{-4}$ yr$^{-1}$. Given $10^5$ galaxies within 200 Mpc, this
amounts to $1.25\times 10^3$ Gpc$^{-3}$ yr$^{-1}$, to be increased
by factor 8 to $10^4$ Gpc$^{-3}$ yr$^{-1}$ by the greater rate of
star formation at the time of binary formation (at $z\sim 1$). Note
that the Nakar et al. ``best guess" includes a factor of 50 for
beaming, so there is a factor of 500 between the ``best guess" and
the Bethe/Brown mergers for LIGO, which would not have a beaming.
Without the beaming the factor would be only 10.

As we shall come back to later, the final Nakar et al. estimate for
initial LIGO if the merging binary is one of two NS's is 0.3 mergers
per year, just 10 times the central Kalogera et al. (2004) value, or
3 mergers per year in the case the the binaries are NS-BH in nature,
the 10 times larger value coming from the large chirp mass (from
$10\msun$ BHs).

We have somewhat different predictions as developed earlier in this
paper. Going back to our factor of $\sim 5$ times more LMBH-BH
mergings than those of binary NS's, we would begin with the Kalogera
et al. 0.003 binary NS mergers in our Galaxy and multiply this
number by factor 8 for the increased rate of star formation at
redshift $z=1$.  Our prediction is about 1 merging per year for LIGO
I, mostly from BH-NS binaries.

In the Nakar et al. (2005) the SHBs are found to result from binaries
with long lifetimes as we discussed in our Sec.~\ref{sec3}. They suggest
that these binaries are either old, invisible double NS or NS-BH binaries.
For the latter case, they chose a BH mass of $\sim 10\msun$,
which would be wonderful for LIGO because it would imply a large chirp
mass and the gravitational waves from the merger of such a BH-NS
binary should be observed rather soon in the LIGO observations.
In the Bethe and Brown (1998) scenario the BHs in these binaries
would be more like $\sim 2\msun$.

A very important new point of the Nakar et al. work is that the invisible
binaries come from a very old population $\sim$ 6 Gyr old;
in other words they were formed when the Universe was only $\sim \frac 12
\tau_{Hubble}$.
The NS in a BH-NS binary will not be recycled so that it will
be observable for $\sim 5\times 10^6$ yrs (see eq.~(\ref{eq4})),
or $\lsim 1$ part in a million of the binary lifetime.

Another important point of Nakar et al. is that if the binaries were born
so long ago (We choose redshift $\sim 1$) then the star formation rate
was substantially higher then.

Nakar et al. (2005) base their redshift distribution model of star formation history on
the Porciani and Madau (2001)
\be
SFR_2(z) \propto
\frac{\exp(3.4 z)}{\exp(3.4 z) +22}
\frac{ [\Omega_m (1+z)^3 + \Omega_\Lambda]^{1/2}}{(1+z)^{3/2}}
\ee
with $\Omega_m=0.3, \Omega_\Lambda =0.7$ and $H_0 = 70$ km$^{-1}$
Mpc$^{-1}$ in standard cosmology. We obtain a factor of 8 higher
star formation at redshift 1 as compared with $z=0$. Whereas such a
large factor should not be applied to the double NS merging rate
because these binaries were not formed so long ago, it should be
applied to the Bethe and Brown (1998) LMBH-NS binaries.

We, of course, believe that we predicted the short-hard bursts. Our
theory would explain the approximate uniformity in them, as arising
from compact objects with essentially NS masses, increased $\sim
0.7\msun$ in the LMBHs by accretion.

None the less we believe the Nakar et al. analysis of the SHBs to be
very useful, in that they establish without model  restrictions that
LIGO should observe an order of magnitude mergings for the binary
compact objects.


\section{Maximum Neutron Star Mass }

Brown and Bethe (1994) claimed, based on the numerical calculations of
Thorsson et al. (1994), that because of kaon condensation which sets
in at densities $\rho\sim 3 n_0$, where $n_0$ is nuclear matter
density, the maximum NS mass is $\sim 1.5\msun$. We discussed
our determination of the gravitational mass of SN1987A, which we believe
went into a BH, in Sec.~\ref{sec2}.
We found it to be $\lsim 1.56\msun$. The maximum NS mass
is decreased by kaon condensation because the electrons, which have Fermi
energies change with kaon condensation into a Bose condensate of zero
momentum kaons.
This substantially decreases the pressure.

There have been many theoretical calculations of kaon condensation in the
past 10 years, none of them getting kaon condensation at as low a
density as $3\rho_0$. However, the driving force in kaon condensation is
the strangeness chiral symmetry breaking parameter $a_3 m_s$ for which
we took results for the Thorsson et al. (1994) value of
$a_3 m_s=-222$ MeV. Lattice gauge calculations have now determined
this parameter to be $-231$ MeV to within a quoted accuracy of 3 to 4\%
(Dong, Lagae, and Liu 1996).
None of them used an $a_3 m_s$ this large in magnitude.

Recently Brown, Lee, Park and Rho (2006) have greatly simplified the
calculation of kaon condensation by calculating about the fixed
point, the density for chiral restoration, in the Harada and
Yamawaki (2003) Vector Manifestation of the hidden local symmetry
theory. The kaon condensation is run completely by the vector meson
degrees of freedom as their masses and coupling constants approach
zero at the fixed point. We move back to the somewhat lower density
for kaon condensation by increasing the Harada and Yamawaki
parameter $a$, which is unity at the fixed point, to $a \lsim 1.3$
obtained by renormalization group analyses. The value of
$\Sigma_{KN}$, and the behavior of strange hyperons are irrelevant
in the new analysis. The Thorsson et al. (1994) result is again
arrived at.

Since Table~\ref{tab1} contains three masses, those of 4U 1700$-$37,
Vela X-1 and J0751$+$1807 which exceed our $1.5\msun$ maximum NS
mass, we should comment briefly.

{\it 4U 1700$-$37} :
Although this compact object has the same accretion history as the
other high-mass X-ray binaries, it doesn't pulse like the others.
Brown, Weingartner and Wijers \cite{BWW} evolve the compact object
as a LMBH.

{\it Vela X-1} : J. van Paradijs et al.~ \cite{paradijs2} pointed
out that in this binary with floppy B-star companion, the apparent
velocity can in some cases increase by up to 30\% (from the
surface elements of the companion swinging around faster than the
center of mass) ``thereby increasing the apparent mass of the
compact object by approximately the same amount". In any case,
Barziv et al.\cite{barziv}  from which the Vela X-1 NS
mass in our table comes, say ``The best value of the mass of Vela
X-1 is $1.86\msun$. Unfortunately, no firm constraints on the
equation of state are possible, since systematic deviations in the
radial-velocity curve do not allow us to exclude a mass around
$1.4\msun$ as found for the other NS's."

{\it J0751$+$1807} :
We consider the measurement of a $2.1\msun$ NS mass in this
NS white dwarf binary a serious challenge to our maximum
NS mass (Nice et al. 2005). It will be clear in our Section~\ref{sec8}
that in the evolution of NS, white dwarf binaries, sufficient mass
is furnished during the red giant evolution of the white dwarf progenitor,
often in conservative mass transfer so that if accepted by the NS,
then most of them would have masses in the vicinity of the quoted mass
in J0751$+$1807 or higher, as found by Tauris and Savonije (1999).
These authors did not introduce the propeller effect, whereas
Francischelli et al. (2002) found that in evolution of double NS
binaries this effect often cuts the accretion down by an order of magnitude.

We have given our reasons earlier \footnote{Bethe and Brown (1995) estimated
the Ni production in 1987A to have come from a NS of maximum mass
$1.56\msun$, whereas these authors believed that the NS had later
evolved into a BH.} that the maximum NS mass cannot be far
above $1.5\msun$.


J0751$+$1807 has a short orbital period of $P_b=6.3$ hours. The short
orbital period allows the detection of the effect of gravitational
radiation emission. According to general relativity the time
rate of change of the orbital period is
\be
\dot P_b &=&
-\frac{192}{5}\left(\frac{P_b}{2\pi}\right)^{-5/3}
\left(1+\frac{73}{24}e^2+\frac{37}{96} e^4\right)
\nonumber\\
&&
\times (1-e^2)^{-7/2} T_\odot^{5/3}
{m_{\rm WD} M_{\rm NS}}{(m_{\rm WD}+M_{\rm NS})^{-1/3}}.
\ee
Since $m_{\rm WD} \ll M_{\rm NS}$, the dependence on masses of $\dot
P_b$ is proportional to $m_{\rm WD} M_{\rm NS}^{2/3}$ or for a given
$\dot P_b$, $M_{\rm NS}\propto m_{\rm WD}^{-3/2}$. Thus, for $m_{\rm
WD} \sim 0.24\msun$, $M_{\rm NS}\sim 1.5\msun$. The NS mass of Nice
et al. (2005) has mass $2.1^{+0.4}_{-0.5}\msun$ at 95\% confidence
level; i.e., at this level $M_{NS}$ could be as low as $1.61 \msun$.


Brown et al. (2006b) have indicated that there is a small correction
to Brown et al. (2006a) who calculated the maximum stable NS mass to
be $1.5\msun$ by fluctuating about the fixed point in the
Harada-Yamawaki (2003) renormalization group formalism. This arises
because the $K^-$-mesons in the kaon condensate which causes the NS
to collapse into a BH have a fermion substructure; i.e., $K^- =|\bar
u s\rangle$ and the $\bar u$ and $s$ quarks are fermions. Thus,
there is a repulsion between $K^-$-mesons when brought together from
the Pauli exclusion principle. The quarks are, however, not very far
from being current quarks, which they become as $n\rightarrow
n_{\chi SR}$, the latter being the chiral symmetry restoration
density, since the critical density for kaon condensation is
\be
n_c \sim \frac 34 n_{\chi SR}.
\ee
Current quarks are thought to be very small in extent.

\begin{table}
\begin{center}
\caption{Calculated accretion onto the pulsar during H and possibly
He red giant stage. $M_i$ is the initial pulsar mass, taken to be
that of the companion, and therefore a lower limit, $\Delta M$ is
the calculated mass accretion onto the first born NS, $M_f$ is the
final pulsar mass following accretion, and $\hat P$ is the
probability of unequal masses of compact objects in NS binaries,
therefore, $1-P$ of Table~\ref{tabX1}. The He core mass of giant
star is assumbed to be $M_{\rm He} = 0.08 (M_{\rm
Giant}/\msun)^{1.45} \msun$. The error in the companion masses less
than $14\msun$ come from our $\sim 0.2\msun$ uncertainty in
accretion in the He red giant evolution if a NS is to remain.}
\label{tabX}

\vskip 4mm
\begin{tabular}{ccccc}
\hline
Giant Mass [$\msun$] & $M_i$ [$\msun$] & $\Delta M$ [$\msun$] & $M_f$ [$\msun$] & $\hat P$ \\
\hline
20 & 1.39 (B1913$+$16)     & 0.87          & 2.26  & 0.92 \\
19 & 1.38                 & 0.84          & 2.22 & 0.92 \\
18 & 1.37                 & 0.82          & 2.19 & 0.91 \\
17 & 1.36                 & 0.79          & 2.15 & 0.90 \\
16 & 1.35                 & 0.76          & 2.11 & 0.89 \\
15 & 1.34 (B1534$+$12)    & 0.74          & 2.08 & 0.88 \\
14 & 1.33                 & 0.71 $-$ 0.91 & $2.04-2.24$ & 0.86 \\
13 & 1.31                 & $0.67-0.87$   & $1.98-2.18$ & 0.83 \\
12 & 1.28                 & $0.63-0.83$   & $1.91-2.11$ & 0.76 \\
11 & 1.25 (J0737$-$3039B) & $0.60-0.80$     & $1.85-2.05$ & 0.56 \\
10 & 1.18 (J1756$-$2251)  & $0.55-0.75$     & $1.73-1.93$ \\
\hline
\end{tabular}
\end{center}

\end{table}

Astronomers are not easily convinced by formal arguments, so we take
the approach of Lee et al. (2006). Using the extensive and accurate
calculation of BKB (2002) of hypercritical accretion, which removed
the approximation of neglecting the compact object mass of Bethe and
Brown (1998), Lee et al. (2006) calculated a table of masses for the
relativistic NS binaries, which we show in Table~\ref{tabX}. $M_f$
is the final pulsar mass after having accreted matter from the giant
companion, while in common envelope with the latter while it is in
red giant stage. The $0.2\msun$ uncertainty for the pulsars from
giants of masses $<15\msun$ comes from accretion in the He star red
giant stage when additional mass can be transferred.

The higher probability $P$ for the more massive giants such as ZAMS
$20\msun$ results because their companion must come from a giant of
lower ZAMS mass, as in Table~\ref{tabX1}. We have given reasons
earlier for our placement in the Table~\ref{tabX} of known pulsars
B1913$+$16, etc.

From the Table~\ref{tabX} we see that the overwhelming probability
would be for the pulsars in B1913$+$16 and B1534$+$12 (and
2127$+$11C if not evolved from NS exchanges in the globular
cluster), J0737$-$3039B and J1756$-$2251 to be much more massive
than the companion. However, if the maximum NS mass is $1.7\msun$ or
less, all of the binaries calculated in the Table~\ref{tabX} with
hypercritical accretion will be BH-NS binaries.

At 95\% confidence, the errors of $^{+0.4}_{-0.5}\msun$ are large,
although they will be made smaller with longer observing time, and
the central value of $2.1\msun$ for J0751$+$1807 is arrived at
through a Bayesian analysis. We feel that it is fair to use the 95\%
confidence limits, since it sticks up well above the other masses in
Table~\ref{tab1}

\section{Hypercritical Accretion in Case BB mass transfer}
\label{sechyper}

We now consider in more detail the two lowest mass binary NS
systems, the double pulsar with mass $1.337 \msun$ for J0737$-$3039A
and $1.250\msun$ for J0737$-$3039B, and J1756$-$2251 with pulsar
mass of $1.40\msun$ and companion mass of $1.18\msun$, which would
have come from $10\msun$ or $11\msun$ ZAMS giants. The He stars in
the He-star, NS binary which precedes the double NS binary will have
been the least massive in the binary NS evolution, so it is
reasonable that they evolve into a He red giant (Dewi and van den
Heuvel 2004). Dewi and van den Heuvel (2004) did not, however,
consider that mass can be accreted onto the NS during the He-star
common envelope evolution. Dewi and van den Heuvel inadvertently
restricted the directions of the supernova kick imparted to PSR
J0737-3039B to the presupernova plane. In the papers of Willems and
Kalogera (2004) and Willems et al. (2005) the investigation was
significantly extended by incorporating proper motion constraints
and the kinematic history of the system in the Galaxy into the
analysis.

Hypercritical accretion requires $\dot M > 10^4 \dot M_{\rm Eddington}$
(Chevalier 1993).
In the NS common envelope evolution in hydrogen red giant,
$\dot M$ is typically $10^8 \dot M_{\rm Eddington}$. The typical
hydrogen envelope has a mass of $\sim 10\msun$. The He star red giant
evolution involves only an $\sim 1\msun$ evolved envelope, which is
sufficient for hypercritical accretion. Using the BKB (2002) equations,
which are now absolutely necessary because the Bethe and Brown
approximation of neglecting the mass of the compact object would now
mean neglecting the pulsar mass as compared with an $\sim 2\msun$
He star mass, we find the following results:

For an initial $M_{{\rm He},i} = 2.5\msun$, final $M_{{\rm He},f} = 1.5\msun$
and $M_{\rm NS} = 1.25 \msun$, the latter being the mass of the
first-born pulsar in the double pulsar system,
we find that the pulsar accretes $0.2\msun$ and the
orbit tightens by a factor of 6.7. For $M_{{\rm He},i} = 2.25\msun$,
$M_{{\rm He},f}=1.25\msun$, the pulsar accretes $0.19\msun$ with
tightening by factor 7.4.
Since we want a first-born pulsar mass of $1.25\msun$,
$2.25\msun < M_{{\rm He},i} < 2.5 \msun$ seems appropriate.\footnote{
$M_{{\rm He},f}$
corrected downward by the greater gravitational binding energy when
it evolves into a NS.}
We are unable to
lower the accreted mass much below $0.20 \msun$, which is twice too
much, but would fit the
difference in mass between pulsar and companion in J1756$-$2251.
It does not seem easy to lower the accretion by kick velocities,
as Fryer \& Kalogera did for 1913$+$16 and 1534$+$12 because
the double NS eccentricity is only 0.09, the smallest
in all double NS's. It is possible that there has been
substantial wind loss, which would diminish the envelope mass
and lower the accreted mass. We would have expected the latter to
be larger than it is.
The point we wish to make is
that the amount of mass accreted by the low-mass pulsars in the He
envelope evolution of their companions is likely to be $\lsim 0.2\msun$.
This is not enough to send the pulsar into a BH.

We believe that we can explain why the pulsar, the first born NS
in the double pulsar and in J1756$-$2251 are substantially
($\lsim 0.2 \msun$) more massive than their companion stars. Certainly
they have not gone through the van den Heuvel and van Paradijs (1993)
scenario, or they would have accreted $\sim 0.5\msun$ from their
hydrogen giant companion.

But of course there must have been a double NS formation
scenario because otherwise the first-born NS would have
accreted $\sim 0.5\msun$ from the hydrogen envelope of its companion
and then later another $\lsim 0.2\msun$ from its He companion as the
latter evolved. Thus, put together, the amount of matter accreted
by the first-born NS would have been comparable to that
accreted from the hydrogen red giant stage of the companion in
the more massive $\gsim 1.5\msun$ first born NS's.
These would turn into BHs.

However, if the BH is formed in the lower mass stars
(ZAMS mass $\lsim 15\msun$) then when the companion evolves into
a He red giant,
the BH will be in a common envelope and the binary
will tighten just as effectively as if it were a NS.
Thus, the double pulsar and J1756$-$2251, which must have formed,
at least in our scenario, through the double He star scenario,
should have an order of magnitude more BH-NS counterparts,
with similar tight orbits.

Thus, the double helium star scenario necessitated if the maximum NS
mass is as small as we say, $1.7\msun$, to produce binary NS
systems, results in an order of magnitude more BH-NS binaries, which
will merge at about the same rate as binary NS's.

We are, however, not finished yet in increasing the prediction for
compact object mergers. To date these are mostly based on the
NS binaries observed in the Galaxy, with extrapolation
to include the number of similar systems that should exist in the
Galaxy.

The observation of the double pulsar increased the estimate by an
order of magnitude.  The radio signal in the double pulsar is much
weaker than that in the Hulse-Taylor 1913$+$16 and it must be
generally true that the radio signals from the pulsars from the
lower mass ZAMS stars are generally much weaker than those from the
pulsars already discovered in binaries. So we may be missing many of
these.

We end this section by referring to yet another channel suggested by
Belczynski and Kalogera (2001). They develop the scenario following
the double He star evolution. For the lower mass stars there is
second stage of common envelope evolution, now in He red giant
envelopes, similar to that for the nearly equal mass evolving
hydrogen stars. The survival probability of the two (Type Ic SN) is
larger, given the tight orbit before the explosions. The end product
is a close NS-NS with very short merger time.

Delgado and Thomas (1981) show that case BB mass transfer is similar
to the double He star scenario for He stars of mass greater than the
Chandrasekhar mass of white dwarfs, in that Case BB mass transfer
can only start after the ignition of carbon, in the sense that helium
core burning has already begun when the hydrogen red giant phase takes
place in Case B mass transfer in the stars massive enough to evolve
into NS's.
However, the chance that the two He stars burn carbon at the same
time is rather small, the ratio of carbon burning time to He burning
time being only $\sim 0.03$ in ZAMS $12\msun$ stars (Schaller et al. 1992).
We thus believe that only $\sim 1\%$ of the binaries will go through
the double common envelope, the others involving Case BB
mass transfer.

BKB (2002) and Belczynski \& Kalogera (2001)
have motivated a large increase in gravitational mergings
by taking the effects of He star red giants; i.e. Case BB mass
transfers, into account. Our more schematic estimates suggest that
overall they were unduly conservative in their estimates.

Be that as it may, we wish to add the order of magnitude more
BH-NS binaries which can be inferred from the
presently measured binary NS masses. In these the
BHs will be more massive than the NS's because
of accretion and their large chirp masses will make them
observable to larger distances.

\section{High Mass X-ray Binaries}
\label{secHMXB}

High mass X-ray binaries give the least reliable information about
NS masses. In general the giant is quite ``floppy" with
various pulsational excitations, the center of mass of the system is inside
of the binary and the NS heats up the giant by radiation.
Over the years the determination of the NS masses have
changed substantially, but with two exceptions 4U 1700$-$37 and
Vela X-1, the NS masses are at an average of $\sim 1.4\msun$
and are consistent with a limit of $<1.5\msun$.

Brown, Weingartner, and Wijers (1996) considered 4U 1700$-$37, which does
not pulse in X-rays and has a harder spectrum than other X-ray binaries.
The accretion history is, however, similar to that of others.
The indication is that the compact object is a BH.

Wellstein and Langer (1999) argue that 4U 1700$-$37, composed of
a $30\msun$ O star with $2.6^{+2.3}_{-1.4}\msun$ compact companion
(Heap and Corcoran 1993, Rubin et al. 1996) must have gone through
a nonconservative mass transfer in late Case B or Case C. In this
way the system loses substantial amounts of mass and angular
momentum and thus becomes a short period binary.
(1700$-$37 has a period of 3.412 days.)

In fact, in Brown et al. (2001a) had the first mass transfer from
the primary to the secondary been case A, AB or B, the primary
of $\sim 30\msun$ would have gone into a NS.\footnote{
Rate of wind loss during He burning
is not sufficiently well known to say that the primary
would have gone into a NS. It might very well have immediately
gone into a LMBH.}
However, with mass transfer from the evolving secondary, it would have
gone into a LMBH.

{}From an evolutionary point of view 4U 1700$-$37 is very
interesting because it is the only binary with low-mass compact
object, albeit BH, for which one can make the argument that it comes
from such a massive region of $> 30\msun$. (Brown et al. 1996
estimate $40\pm 10\msun$.)

Vela X-1 may well be the worst system in which to measure the mass
of a NS.

``Another systematic effect, due to the distortion of the primary may be
quite important in the case of X-ray binaries with a small mass ratio\footnote{
Authors: Vela X-1 is made up of an $18\msun$ B-star and NS.
The center of mass is well within the giant.}
as the Vela X-1 system. In such a system the radial velocities of certain
individual surface elements of the primary are much greater than the
orbital velocity of the center of mass of this star, in the case of
synchronous rotation. When the primary is tidally distorted and has
a variation of effective temperature and gravity across its surface,
it is by no means clear that the observed radial velocity, which is given
by some spectrophotometric average over the surface, can be identified
with the motion of the center of mass of the object."
(J. van Paradijs et al. 1977a)

``In a previous paper we presented a numerical study of the effect of the
deformation of the primary on its apparent radial velocity
(van Paradijs et al. 1977a). The apparent velocity amplitude can in some
cases increase by up to 30 \%, thereby increasing the apparent mass of the
compact object by approximately the same amount."
(J. van Paradijs et al. 1977b)

It is known that the light curve varies substantially from night to night.
Indeed, in Barziv et al. (2001) from which the large NS mass
is taken, the authors say ``The best estimate of the mass of Vela X-1 is
$1.86\msun$. Unfortunately, no firm constraints on the equation of state
are possible, since systematic deviations in the radial-velocity
curve do not allow us to exclude a mass around $1.4\msun$ as found for
other neutron stars."

\section{Carbon-Oxygen White-Dwarf, Neutron Star Binaries}
\label{sec6}

The high-field eccentric binaries B2303$+$46, long thought to be
a wide NS binary and J1141$-$65 have not gone through
common envelope evolution which would have circularized them
and brought their magnetic fields down. The magnetic field of
B2303$+$46 is $7.9\times 10^{11}$ G, that of J1141$-$65,
$1.3\times 10^{12}$ G. The relative time that such a binary can be
seen, before it goes into the ``graveyard" goes inversely with $B$.
Thus we have ``observability premium", Eq.~(\ref{eq5}),
equal to essentially unity as an average for the two unrecycled
binaries above.

Tauris et al. (2000) proposed five binaries which they evolved through
common envelope. Two of these recycled pulsars in relativistic orbits
PSR 1157$-$5112 and J1757$-$5322 are discussed by Edwards and Bailes
(2001). Two others J1435$-$6100 and J1454$-$5846 are discussed by
Camilo et al. (2001).

The fifth of the systems favored for common envelope evolution by
Tauris, van den Heuvel and Savonije (2000) is J1022$+$1001. This closely
resembles PSR J2145$-$0750, aside from a more massive white dwarf
companion, as remarked by van den Heuvel (1994a), who evolved
the latter through common envelope when the white dwarf progenitor
was on the AGB (Case C mass transfer). Van den Heuvel suggested for
J2145$-$0750 that there was considerable mass loss because of possible
instabilities on the AGB caused by the presence of the NS.
This is one possibility of saving our general theme; i.e., that most
of the NS, carbon-oxygen white dwarf binaries would end up
as LMBH carbon-oxygen white dwarf binaries, although
some might be saved with NS's because of the possible
instabilities caused by the NS while the white dwarf
progenitor is on the AGB.
Van den
Heuvel chose $\lambda=1/2$ for the parameter that characterizes
the structure of the hydrogen envelope of the massive star that is
removed in common envelope evolution. Dewi \& Tauris (2001)
have since carried out detailed calculations that in some cases,
``particularly on the asymptotic giant branch of lower-mass
stars, it is possible that $\lambda>4$." This lowers the
binding energy of the envelope by a large factor, so that
it can be removed in common envelope evolution and still leave a
reasonably wide orbit, as remarked by Dewi \& Tauris. We believe
that this may be the reason that some binaries have survived
common envelope evolution.

\begin{table}
\caption{Inferred magnetic fields $B$ and the observability
premium $\Pi$ for recycled pulsars}
\label{tabPS}
\begin{center}
\begin{tabular}{ccc}
\hline
Pulsars & $B$ & $\Pi$ \\
\hline
J2145$+$0750  &  $6\times 10^8$ G & 1667 \\
J1022$+$1001  &  $8.4\times 10^8$ G & 1190 \\
J1157$-$5112  &  $<6.3 \times 10^8$ G & $>$ 159 \\
J1453$-$58    & $6.1\times 10^9$ G & 164 \\
J1435$-$60    &  $4.7\times 10^8$ G & 2127 \\
\hline
\end{tabular}
\end{center}
\end{table}

In four of the six recycled pulsars (assumed to have been evolved
through common envelope evolution) the magnetic field have been
inferred as in Table~\ref{tabPS}. The observability premium $\Pi$
is high for most of these pulsars. Since we see two unrecycled
pulsars with high magnetic fields $B\sim 10^{12}$ G;
therefore, $\Pi\sim 1$, we should see $\sim 20,000$ recycled pulsars.

The above argument does not take into account the greater difficulty
of observing pulsars with low magnetic fields $\sim 10^8 - 10^9$ G,
which may remove some of the large predicted numbers of recycled
pulsars in case they did not go into BHs during common
envelope evolution.

On the other hand, the NS in these cases goes through
common envelope with a star of main sequence less than $\sim 10\msun$
since it must end up as a white dwarf, albeit a relatively massive
carbon-oxygen one. As discussed by van den Heuvel (1994a) only part
of the energy to remove the envelope will be connected with the
accretion, the rest coming from wind losses, etc. Thus, our
observation that most of these binaries must end up as white-dwarf,
low-mass X-ray binaries gives credence to essentially all of the
NS's which go through common envelope evolution in the
evolution of binary NS's ending up as LMBHs.

As noted earlier, there are hopes that during our lifetime - at
least that of one or two of the authors - this can be tested, because
the Bethe-Brown prediction of the factor 20 greater contribution of
NS-LMBH to binary NS mergers should
be robust, essentially independent of the number of the latter.

\section{White Dwarf-Neutron Star Binaries}
\label{sec8}

This class of 12 (See Table~\ref{tab1}) is the most numerous class.
They mostly would be expected to come from a NS with main sequence
star of mass between $1\msun$ and $2\msun$, the $1\msun$ because
they must evolve in a Hubble time. The main sequence star evolves,
either transferring matter to the NS or matter is lost by wind,
because it ends up as a typically quite low-mass white draft.

\def\espace{\phantom{xxx}}

\begin{table}
\caption{White Dwarf Companion masses ($m_2$) and orbital period ($P$)
in NS, He-White-Dwarf Binaries.
Refs; 1) Thorsett \& Chakrabarty 1999,
2) Hansen \& Phinney 1997,
3) Tauris, van den Heuvel, \& Savonije 2000,
4) Lundgren, Zepka, \& Cordes 1995,
5) Navarro et al. 1995,
6) Thorsett, Arzoumanian, \& Taylor 1993,
7) van Kerkwijk et al. 2000,
8) Lyne et al. 1990,
9) Phinney \& Kulkarni 1994.
}
\label{tab2}
$$
\begin{array}{lccl}
\hline
\rm Pulsar &\rm  m_2\ (\msun) &\rm\espace P\ (days) \espace
&\rm References\\
\hline
J0034-0534              &   0.15 - 0.32  & 1.589   &  1,2\\
J0218+4232              & 0.2            & 2.029   & 1,5 \\
J0751+1807              &  0.15          & 0.263   & 1,2,4 \\
J1012+5307              & 0.165 - 0.215  & 0.605   &  1,2\\
J1045-4509              & < 0.168        & 4.084   &  1\\
J1232-6501              &  0.175         & 1.863   & 3\\
J1713+0747              & 0.15 - 0.31    & 67.825  &  1,2\\
B1744-24A \ (J1748-2446A) & 0.15         & 0.076   & 1,8 \\
B1800-27 \ (J1803-2712)  & 0.17          & 406.781 & 1,9 \\
J1804-2718              &  0.185 - 0.253 & 11.129  &  1\\
B1855+09 \ (J1857+0943)   &  0.19 - 0.26   & 12.327  &  1,2\\
J2129-5721              &   0.176        & 6.625   &  1\\
J2317+1439                & 0.21         & 2.459   & 1,9 \\
\hline
J0437-4715              &  0.15 - 0.375  & 5.7     &  1,2\\
J1455-3330              &   0.305        & 76.174  &  1\\
B1620-26 \ (J1623-2631) &  0.3           & 191.443 & 1,6 \\
J1640+2224              &   0.25 - 0.45  & 175.460 &  1,2\\
J1643-1224              &   0.341        & 147.017 &  1\\
B1718-19 \ (J1721-1936) &  0.3           & 0.258   & 1,7 \\
B1802-07 \ (J1804-0735)   & > 0.29         & 2.617   &  1\\
J1904+04                &  0.27          & 15.75   & 3\\
B1953+29 \ (J1955+2908)   &   0.328        & 117.349 &  1\\
J2019+2425              & 0.264-0.354    & 76.512  &  1,2\\
J2033+17                &   0.290        & 56.2    &  1\\
J2229+2643              &   0.315        & 93.015  &  1\\
\hline
\end{array}
$$
\end{table}

In Tab.~\ref{tab2} we have collected the helium white dwarf masses we could
find. Note that in particular the binaries B2303$+$46 and
J1141$-$6545, which we discussed in Sec.~\ref{sec6}, do not
appear in our table. They have quite massive carbon-oxygen white
dwarf companions, and were discussed in the last section.

\begin{table}
\caption{Statistics of the mass distribution of white dwarfs.
         Logarithmic distribution of the initial NS,
         main sequence star is assumed. Last column is the
         number of observed systems summarized in Table~\ref{tab2}.}
\label{tab3}
$$
\begin{array}{cccc}
\hline
\espace m_2 \ (\msun) \espace &\espace  R_g \ (\rsun)\espace  &\espace   \ln (R_u/R_l)\espace  &\espace  \rm Observations\\
\hline
0.15-0.25 & 1.47-10.0 & 1.92 & 10 \\
0.25-0.35 & 10.0-42.0 & 1.44 & 12 \\
0.35-0.46 & 42.0-128  & 1.11 &  0 \\
\hline
\end{array}
$$
\end{table}

In Tab.~\ref{tab3} we show the statistics of the mass distribution of white
dwarfs. We note that all of our tabulated white dwarfs have masses
$\lsim 0.35\msun$, whereas single white dwarf tend to peak up at
$\sim 0.6\msun$. This indicates to us immediately that the companion
NS is strongly influential in increasing the wind loss.
(See the suggestion of van den Heuvel about J1245$-$0750 in the
last section that there was considerable mass loss because of possible
instabilities on the AGB (asymptotic giant branch) caused
by the presence of the NS.)

The white dwarf, NS binaries have been evolved by
Tauris and Savonije (1999). For evolution with stable
mass transfer, i.e., for main sequence masses less than $M_{\rm NS}$,
the evolution was basically conservative, matter accreting below
or at the Eddington limit, onto the accretion disc of the white
dwarf. In the case of the main sequence mass
$M_{\rm MS} > M_{\rm NS}$, for $M_{\rm MS}$ up to $2\msun$,
the evolution was still conservative from the standpoint of the
accretion disk, but the amount above the Eddington limit for
the white dwarf was expelled with the angular velocity of the
NS.

The mass distribution of NS's of Tauris and Savonije do not give
NS masses that look anything like those shown in Table~\ref{tab1}.
Even though Tauris and Savonije began with the somewhat small NS
mass of $1.3\msun$, they have copious numbers up to $2\msun$.
Only J0751$+$1807 in Tab.~\ref{tab1} really comes this high. We believe
the reason for their high NS masses is explained in the last
sentence of the caption to their Fig.~4:
``The post-accretion $M_{\rm NS}$ curves (bottom) assume no mass
loss from accretion disk instabilities of propeller effects."
As we discussed in Sec.~\ref{sec3}, Francischelli et al. (2002)
found that the propeller effects decreased the mass accretion
from He star wind onto the pulsar by an order of magnitude.
In other words, the accretion may be scaled down from the
$\sim 1\msun$ difference between progenitor main sequence
and white dwarf masses, to $\sim 0.1\msun$
actually accepted, the remainder being lost through the agency
of the propeller effect. Of course, because of different angular
momenta in the different binaries we can only give order of
magnitude estimates.

Since Brown and Bethe (1994) ``A scenario for a large number of
low-mass black holes in the Galaxy",
at which time NS masses were spread rather widely,
we have got used to seeing their masses fall below our projected
maximum $1.5\msun$. Thus, at that time, the NS in
J2019$+$2425 had an upper limit on its mass of $\sim 1.64\msun$
from the white dwarf mass-period relation.
Nice et al. (2001) could constrain the inclination angle of the
binary to $i<72^\circ$ from the proper motion of the binary,
with a median likelihood value of $63^\circ$. A similar
limit on inclination angle arose from the lack of a detestable
Shapiro delay signal. The NS mass was determined to be
at most $1.51\msun$.

\section{Summary Conclusion}

Our Selected Papers with Commentary is entitled ``Formation and
Evolution of Black Holes in the Galaxy". We were shifted from
NSs to BHs by SN 1987A which showed that
a relatively low-mass compact core could evolve into a LMBH.

Most important is our double helium star scenario for the evolution
of binary NS's. The motivation of this began with Chevalier (1993)
who estimated that a NS would accrete $\sim 1\msun$ in common
envelope evolution. We confirmed and made more quantitative his
estimate. With addition of $0.75\msun$, the NS will certainly go
into a BH. The double helium star scenario avoids the NS having to
go through common envelope evolution.

The near equality of the masses of pulsar and companion in the
binary NS's is observational support for our double NS scenario.
Where the masses of pulsar and companion are substantially
different, in the Hulse-Taylor binary, this difference tells us the
range of ZAMS masses that the binary came from. In this case, from
the most massive possible range $\sim 20\msun$ where the Fe core
masses change most rapidly with main sequence mass, because the
companion mass is almost 4\% lower than that of the pulsar.

The importance of our double He star scenario is that one must then
look for the fate of the NS's which do go through common envelope
evolution in the envelope of the evolving giant. Bethe and Brown
(1998) show that these do, indeed, produce LMBH fresh NS binaries.
There are a factor of 5 more of these than double NS binaries and
because of the higher mass of the BH, they give a larger factor in
expected mergings.

While the recently discovered double pulsar is very interesting (and
was very improbable to be observed) it simply brings the binary NS
contribution to LIGO up to snuff, and does not change our additional
factor of 5, although it does greatly increase the merging rate of
the binaries one sees.

We believe that the many recently observed short hard gamma-ray
bursts give credence to the large number of LMBH-NS binaries we
predicted and even our factor 40, which includes the higher star
formation rate in the early Universe, is easily subsumed in the
number of SHBs. However, the gravitational waves from these mergings
may still not be sufficient for LIGO observations in the next few
years. What LIGO needs is a merger with large chirp mass, as
predicted by Portegies Zwart and McMillan (2000) which overwhelms
the background. Of course in time, possibly a few years, the LIGO
sensitivities should observe the merger of the lower chirp mass
binaries discussed here.

\section{Discussion}

Kip Thorne asked Hans Bethe and Gerry Brown
to work out the mergings of NS-BH binaries while they were in Caltech in 1996.
This was a new activity for Hans, who was 90 at the time, and he attacked it
with gusto.
From the paper of Brown (1995) it was clear that there were an order of
magnitude more of these than of binary NS's because the standard
scenario for making the latter always ended up with the first
NS going into a BH during its common envelope with the
companion while in red giant stage.

The authors returned to this problem in 2003, after publishing
their joint works (Bethe, Brown and Lee 2003) and Hans was engaged with this
problem right until his death. In fact,  he had a discussion of it
on the telephone with Gerry Brown the morning of the day of his death.

Crucial to our work on evolving the binary objects was Hans' analytic
common envelope evolution. This is reproduced in his
obituary by Brown (2005). It was carried through with Kepler's and Newton's
Laws using elementary calculus.

\section*{Acknowledgments}
This is written in celebration of Hans Bethe's 100th birthday. We
would like to thank Marten van Kerkwijk and Ralph Wijers for many
helpful discussions. We wish to especially than Ed van den Heuvel,
who has always stimulated us, and to our scenarios which sometimes
contradict his earlier ones, says ``New ideas are always welcome".
GEB was supported in part by the US Department of Energy under Grant
No. DE-FG02-88ER40388. CHL  was supported by the Korea Research
Foundation Grant funded by the Korean Government(MOEHRD, Basic
Research Promotion Fund) (KRF-2005-070-C00034).



\renewcommand{\thesection}{Appendix~\Alph{section}}



\end{document}